\documentclass{llncs}

\usepackage{graphicx}

\usepackage{amsmath,amssymb,amsfonts,textcomp,mathtools}

\usepackage{authblk}

\usepackage{url}

\usepackage{color}

\begin{document}

\title{A German Corpus for Text Similarity \\Detection Tasks}

\author{
Juan-Manuel Torres-Moreno \inst{1,2}  
\and 
Gerardo Sierra \inst{1,3}
\and
Peter Peinl \inst{4}
}

\institute{
Universit\'e d'Avignon et des Pays de Vaucluse\\
Laboratoire Informatique d'Avignon\\
\footnotesize{juan-manuel.torres@univ-avignon.fr}\\
\and
\'Ecole Polytechnique de Montr\'eal\\
Departement de G\'enie Informatique\\
\and
Universidad Nacional Aut\'onoma de M\'exico\\
Instituto de Ingenier\'ia\\
\footnotesize{gsierram@iingen.unam.mx}\\
\and 
University of Applied Sciences Fulda\\
Faculty of Applied Informatics \\ 
\footnotesize{peter.peinl@informatik.hs-fulda.de}\\
}

\vspace{2cm}

\maketitle

\begin{abstract}
Text similarity detection aims at measuring the degree of similarity between a pair of texts. 
Corpora available for text similarity detection are designed to evaluate the algorithms to assess the paraphrase level among documents. 
In this paper we present a textual German corpus for similarity detection.
The purpose of this corpus is to automatically assess the similarity between a pair of texts and to evaluate different similarity measures, both for whole documents or for individual sentences. 
Therefore we have calculated several simple measures on our corpus based on a library of similarity functions.\footnote{Preprint of \textsl{International Journal of Computational Linguistics and Applications}, vol. 5, no. 2, 2014, pp. 9--24}
\end{abstract}


\section{Introduction}
\label{intro}

Text similarity is a condition or property that can be measured between two or more texts, which determines the degree of similarity between them.
Text similarity ranges between 0\% (no relationship at all) and 100\% (documents are identical).
Also note that two similar texts do not need to share the content, neither verbatim nor expressed in other words.
They may just cover the same topic or merely be written in the same language.

Similarity detection has been intensively studied and is of great interest for different applications of Natural Language Processing (NLP), such as plagiarism and paraphrasing detection, fraud analysis, document clustering, machine translation, automatic text summarization and information retrieval.

To develop systems for similarity detection both a training and a test corpus, built to the requirements of the task to achieve, have to be available.
For paraphrase detection, the corpus in particular must comprise the text source and the text paraphrasing the content of the text source.
Corpora specifically designed for this task already exist, such as the METER Corpus\footnote{\url{http://nlp.shef.ac.uk/meter/}}, the Microsoft Research Paraphrase Corpus\footnote{\url{http://research.microsoft.com/en-us/downloads/607d14d9-20cd-47e3-85bc-a2f65cd28042/}} and the PAN Plagiarism Corpus\footnote{\url{http://www.webis.de/research/corpora}}.
However, as they do not fulfil the requirements of a tool we are still working on, we needed an \textit{ad hoc} corpus.

We are aiming at the assessment of the similarity between a pair of documents, not necessarily paraphrase detection.
For that purpose we needed a gold-standard comparable corpus containing source texts and texts similar to them, either because they are paraphrases or because they just deal with the same topic.
Even more, because they share the lexical units although do not share the topic\footnote{That issue will not be presented in this paper.}.
Also we aim at a more precise assessment of similarity and at a mapping between the source text and the paraphrased text at the paragraph level.

The purpose of the paper is to present the methodology of the construction of a paraphrasing corpus, the description of the German corpus using this methodology, and an illustration of its usefulness with respect to standard simple measures for paraphrasing detection.

The paper is organized as follows. 
In Section \ref{sim_detection} we give an overview of similarity detection and the current corpora for similarity detection.
Then, in Section \ref{measuring} we outline the usual simple measures to detect and evaluate similarity.
Next, in Section \ref{building} we describe the methodology to build our corpus.
In Section \ref{evaluation} and Section \ref{exploitation} we report on the application and exploitation of different simple measures on our corpus, before concluding in Section \ref{conclusions}.

\section{Similarity Detection}
\label{sim_detection}

Similarity as a concept has a wide range of applications in different areas.
Similarity implies different features and relationships among objects.
Depending on the area, context or perspective, similarity between objects can differ.
Similarity in fact depends on the context surrounding the object.
An object $a$ is similar to $b$ only referring to a context $c$ \cite{goldstone94}.
Every task involving similarity must therefore specify the context and the features to focus on.
For example, two books on the same library shelf are likely to be similar due to the same thematic, even if the content or the language is different.

Hence, similarity detection aims at comparing different objects and to observe the common features they share according to certain parameters.
The units of language to compare in the context of NLP might be words, sentences, paragraphs or documents.

Text similarity ranges between the paraphrase of a sentence or paragraph from another document and a complete copy of a document.
As \cite{bendersky09} explain, there is a similarity spectrum from plagiarism (nearly identical documents or even identical documents) to topical similarity, passing through text reuse.
In addition, two documents may be similar without any direct relationship, but by their similarity to a third text.
For example, different newswires derived from a common source text provided by a news agency are similar, as are the homeworks of pupils on a common thematic or reviews or adaptations of a literary work.

\subsection{Current corpora for similarity detection}
\label{current}

Most of the well-known corpora on similar texts are designed to evaluate the algorithms to detect paraphrases among documents.
Some comparable corpora of considerable size have been created automatically by using certain heuristics.

The METER corpus is a corpus of news texts collected manually from a news agency and nine daily newspapers \cite{clough02,gaizauskas01} that reuse these newswires.
The resulting 1717 texts were manually classified at the document level into three categories, according to the relatedness of the texts to the original newswire: wholly derived (WD), if the note derives fully from the agency; partially derived (PD), if the article uses other sources besides the information provided by the agency; and non-derived (ND), if the note is written independently from the newswire provided by the agency.
At the phrasal level, individual words and phrases were compared to find verbatim text, paraphrased text or none at all.

The Microsoft Research Paraphrase Corpus consists of 5801 pairwise aligned sentences that exhibit lexical and/or structural paraphrase alternations extracted from news reports \cite{dolan04}.
It was created automatically using string edit distance and discourse-based heuristic extraction techniques.
The corpus has binary statements indicating whether human evaluators considered the pair of sentences to be semantically equivalent or not \cite{dolan05}.

The PAN Plagiarism Corpus is a corpus for the evaluation of automatic plagiarism detection algorithms.
For the source documents, texts of artificial plagiarism were created automatically through a heuristic of changing some parameter, such as document length, suspicious-to-source ratio, plagiarism percentage and plagiarism length, plagiarism languages and plagiarism obfuscation \cite{potthast09}.

\section{Measuring text similarity}
\label{measuring}

Metrics for similarity detection assess either the commonality or the difference between two sets of data.
The higher the commonality between two objects, the more similar they are.
On the other hand, the higher the difference between two objects, the lower is their similarity.
Hence, similarity increases with commonality, but decreases with difference \cite{lin98}.

The metrics calculate a score that can be normalized to be between zero and one.
The ranking score is useful for different tasks, such as information retrieval,  Question-Answering (Q\&A) or Automatic Text Summarization systems \cite{torres2014}.
However, for paraphrase detection purposes, a binary result is considered \cite{bar12}, but the similarity measures get a grade as result.
Based on a threshold it is determined whether the compared texts are the same, a paraphrase or different \cite{ali11}.

We can differentiate three main approaches for similarity detection.
They are based either on vector space models (term-based), on text alignment (linguistic knowledge-based) or on n-gram overlapping (string-based).

\subsection{Vector Space Models}
\label{VSM}

Vector Space Models are one of the simplest and most common way to assess content similarity among documents, which are considered as a bag of words.
Therefore, words are supposed to appear independently while the order is irrelevant.
A text is transformed into a term vector representation, following the removal of stop words and stemming.
We focus on three metrics to determine the commonality between two texts (Cosine similarity, Dice similarity and Jaccard similarity), as well as on two metrics to measure dissimilarity (Euclidean distance and Manhattan distance) \cite{gomaa13}.

Cosine similarity is one of the most popular vector based similarity measures.
A text is transformed into a vector space, so that the Euclidean cosine rule can be used to determine similarity.
Cosine similarity between documents $D_1$ and $D_2$ in a vector space is defined as:

\begin{equation}
sim_{C} (\overrightarrow{D_1},\overrightarrow{D_2}) = (\overrightarrow{D_1},\overrightarrow{D_2}) = \sum_{j=1}^{k} w_{1,j} w_{2,j}
\end{equation}
\noindent The $w_{x,y}$ are the weight of the words calculated as the term frequency tf and
$k$ corresponds to the number of different terms.

Dice similarity uses the Dice coefficient, i.e. the ratio of twice the number of shared terms in the compared texts to the total number of terms in both texts.
$m_c$ is the number of common words in documents $D_1$ and $D_2$, and $m_1$ and $m_2$ the number of words of $D_1$ and $D_2$, respectively, Dice similarity is:

\begin{equation}
sim_{D} (\overrightarrow{D_1},\overrightarrow{D_2}) =  \frac{2m_c}{{m_1+m_2}}
\end{equation}

Jaccard similarity measures similarity by comparing the number of common terms to the number of all unique terms in both texts.
$m_c$ being the number of common words between documents $D_1$ and $D_2$, and $m_1$ and $m_2$ the number of words of $D_1$ and $D_2$, respectively, Jaccard similarity is:

\begin{equation}
sim_{J} (\overrightarrow{D_1},\overrightarrow{D_2}) =  \frac{m_c}{{m_1+m_2-m_c}}
\end{equation}

Euclidean distance is an ordinary measure in the vector space model to determine the distance between the vector inputs, rather than the angle as in the cosine rule.
Euclidean distance is defined as:

\begin{equation}
dist_{E} (\overrightarrow{D_1},\overrightarrow{D_2}) = \lVert \overrightarrow{D_1},\overrightarrow{D_2} \rVert = \sqrt{\sum_{j=1}^{k} (w_{1,j}-w_{2,j})^2}
\end{equation}

Manhattan distance can be described in two dimensions with discrete-valued vectors, where the distance value is simply the sum of the differences of their corresponding vectors.

\begin{equation}
dist_{M} (\overrightarrow{D_1},\overrightarrow{D_2}) = \sum_{j=1}^{k} (w_{1,j}-w_{2,j})
\end{equation}

\subsection{Text alignment}
\label{alignment}

Unlike vector space model metrics, text alignment algorithms compare two strings of characters by calculating the number of operations (either on single characters or on words) to transform one string into the other.
Since a dependency exists among the characters/words their order in the text is relevant. 
Depending on the number of operations several algorithms have been defined \cite{ali11}.
We focus on two, Levenshtein distance and Jaro-Winkler distance.

Jaro-Winkler distance is an extension of the Jaro distance metric which takes typical spelling deviations into account.
This extension modifies the weights of poorly matching pairs that share a common prefix.

Levenshtein distance is a simple edit distance function which calculates the distance by simply counting the minimum number of operations needed to transform one string into the other.

\subsection{$n$-gram overlapping}
\label{overlapping}

A common language-independent algorithm used in different NLP tasks is character or word $n$-grams overlapping.
An $n$-gram is a subsequence of n characters or words of a given sequence of text.
For similarity detection, $n$-grams overlapping measures the number of shared words $n$-grams between two texts \cite{clough02}.
The similarity measure is calculated using any appropriate similarity metric, such as Dice or Euclidian.
The simplest way is by dividing the number of similar $n$-grams by the maximum number of $n$-grams \cite{gomaa13}.

A variation is the $k$-skip-$n$-grams overlapping that uses an $n$-grams distance metric, but takes into account a skip of $k$ characters or words.
Therefore, the characters or words need not be consecutive, there may be gaps in between \cite{guthrie06}.
This kind of $n$-grams allows to obtain nonconsecutive textual segments.
The Rouge-$n$ formula between two documents is:

\begin{equation}
\textrm{Rouge-}n = \frac{\sum_{n-grams} \epsilon \textrm{Can} \bigcap \textrm{Ref} \}}{\sum_{n-grams} \epsilon \textrm{Ref}}
\end{equation}

Can are the $n$-grams corresponding to the first document and Ref corresponds to the $n$-grams of the second document.

\section{Building a corpus}
\label{building}

The general purpose of our corpus is to automatically assess the similarity between a pair of texts.
Therefore we evaluate different similarity measures, either on the document or on the sentence level.
That is, we try not only to find out whether two documents are similar, but also which sentences match.

Our corpus contains paraphrases of different complexity levels, from basic (for example by using synonyms) up to more complex ones.
The difference to existing corpora lies in the granularity, in the ranking of paraphrase and in the annotation method for evaluation purposes.
Our granularity is phrase-to-phrase.
Furthermore every phrase is (to a different degree) modified as compared to the source phrase.
So, the whole document is paraphrased.
Related to the ranking, for the source text we obtain several levels of paraphrase.
The first one relates to the bottom level, the second to an upper level and progressively up to the top level.
Finally, we annotate the phrases of the source text mapping with the paraphrased document.

\subsection{Subject and structure of the corpus}
\label{structure}

At the beginning it was decided to base the experiment on an article in German on Wikipedia\footnote{\url{http://de.wikipedia.org/wiki/Baumkuchen}} and build the corpus around the subject of the article.
To limit the amount of paraphrasing work the article should consist of approximately 30 phrases.
A particular cake (\textsl{Baumkuchen}\footnote{For the English version see also \url{http://en.wikipedia.org/wiki/Baumkuchen}}), very well known in Germany, was chosen as the subject of the study.
As the original article contained slightly more phrases than required, a small number (less than 10) of phrases were deleted to obtain the version (31 phrases) used in our experiment.

To achieve the goals of the experiment, the corpus was partitioned into three subsets of documents, all to be evaluated for their similarity to the Wikipedia document.
By systematically applying the rules explained below to the original article, two sets were obtained, each containing 5 manually rewritten (modified) documents.
These sub-corpora are called \textit{basic} and \textit{complex paraphrase}.

The third sub-corpus was constituted for control purposes.
It consists of 10 documents found on the WWW by a careful manual search.
All documents were selected after thorough evaluation.
The author read them all to make sure that the subject (the cake) was adequately addressed in the article.
The documents of this control corpus were further divided into two categories, of 5 documents each.
The first category comprises minimally modified versions of documents that had been cited in the original Wikipedia article.
Consequently, the degree of similarity to the original should be very high.

None of the documents of the second category had been cited by Wikipedia, i.e. their similarity to the original ought to be much lower.
Those documents were also found on the web, but they treat the subject (the cake) from (completely) different angles as compared to the original article, i.e. an interview with the general manager of the company that makes and sells the cake in Japan, where it has made its very successful entry in the 1950ies, an article from a German women's magazine, one that proposes a recipe that does without eggs, etc.

\subsection{Rules concerning form and structure of paraphrased documents}
\label{rules}

The manual rewriting process to obtain paraphrased versions of the original article was guided by a set of rules mainly specifying the permitted alterations to the structure and syntax of the source article.
However, these rules had to be applied sensitively such that the narrative of the resulting article remained cohesive and comprehensible for a human reader.

Basic paraphrase almost ruled out a change of the length of an article, i.e. no more than one phrase was to be added to or deleted from the original.
Equally, exchanges of segments (sub-phrases) among different phrases of the original article were forbidden.
However, segments within a phrase might be arranged in a different sequence (intra-phrase), including the elimination of sub-phrases.
The order of phrases in the paraphrased version of the article might also be a permutation of the original article.
As the article focused on four main aspects of the cake, i.e. history, recipe, production process and geographical reach, the number of semantically acceptable permutations was rather limited by the requirement that the paraphrased version had to be comprehensive and cohesive.

Complex paraphrase gave more leeway to the rewriting process by permitting the insertion of up to 5 new phrases into the document plus the deletion of up to 5 phrases.
In addition, exchanging segments between (several) phrases of the original article was allowed and encouraged.
That is, complex paraphrase both makes use of inter-phrase and intra-phrase exchange of segments (sub-phrases).
The term exchange was defined in a very general way. It encompasses splitting one phrase into two phrases or merging two phrases into one.

Another rule stipulated that none of the phrases of the original document was allowed to appear unaltered in any of the manually paraphrased versions.
Yet small changes to the phrase (in the source document) to be paraphrased, like the removal of an adjective or an element within the enumeration of several alternatives, were sufficient to make the paraphrased document comply with that particular rule.

\subsection{Paraphrasing the meaning of the text}
\label{paraphrasing}

There were no ``semantic'' rules attached to modifications of the original document, as long as the meaning (narrative/content) of the resulting document was more or less equivalent to the original.
This allowed for using more general or more specific terms, the omission of details, the use of synonyms, different representations of information, etc.
Several known standard techniques and tricks were applied and novel ideas developed as the author became more sophisticated in the process of generating further variants of paraphrased documents.
Documents edited at a later time typically made use of knowledge acquired in all previous steps, unless the level of sophistication was deliberately reset or degraded.
 
The following paragraphs give an overview of all the techniques used in the experiment.
The complex ones are generally found in documents of the complex paraphrase corpus.
Also note that the following examples are represented in English, with an as faithful translation as possible.
Among the simpler techniques, a few well known modifications that work for most of the languages shall be mentioned:

\begin{itemize}
 \item \underline{Abbreviations}: ``vs'' or ``versus''.
 \item \underline{Numbers}: can be written as a sequence of ciphers (15) or as text (fifteen), small numbers also in Roman style (XV).
 \item \underline{Enumerations (reordering and suppression)}: ``nutmeg, cinnamon and cardamom'' vs. ``cardamom and nutmeg''.
 \item \underline{Hyperonyms and hyponyms}: ``sugary substance'' vs. ``honey'' vs. ``bee honey'', ``wood'' vs. ``pinewood''.
 \item \underline{Synonyms and definitions}: ``manuscript'' vs. ``handwritten document''. 
\end{itemize}

More sophisticated modifications, some due to the intricacy of the German language, were:

\begin{itemize}
 \item \underline{Compound words}: one specificity of the German language is the extensive use of compound words of very often considerable length. Where in English ``production of cake'' is the proper term, in German ``Produktion von Kuchen'' or ``Kuchenproduktion'' are synonymous, with the latter one stylistically preferable. Further elaborating on the example ``Kuchenproduktionsverfahren'' in English requires at least twice the separator ``of''. In German there are several rewritings, which all may be used as a paraphrase.
 \item \underline{Complex phrase structure}: another specificity of the German language is a certain tendency to formulate lengthy phrases of complex structure, i.e. containing (several layers of nested) several sub-phrases. There typically are many simpler rewritings or the possibility to reorder those sub-phrases without changing the meaning.
\end{itemize}

Complex ``semantically'' paraphrase was achieved by generalizing temporal and geographical entities, indirect definitions of persons that cannot be deduced from other phrases of the original text.
Approximations of quantities have also been used regularly:

\begin{itemize}
 \item \underline{Temporal}: 1855 vs. ``in the midst of the nineteenth century'' vs. ``between 1840 and eighteen hundred and sixty-two''.
 \item \underline{Geographical}: ``Dresden'' vs. ``the capital of Saxony'', ``Japan'' vs. ``the land of the rising sun'', ``Masuria'' vs. ``north-eastern region of Poland''.
 \item \underline{Personal}: ``Prince Elector Frederick William'' vs. ``the head of state of Brandenburg'', ``Karl Joseph Wilhelm Juchheim'' vs. ``German patissier''.
 \item \underline{Quantitative}: 8 vs. ``between 6 and 9'' vs. ``a one digit number''.
\end{itemize}

As the paraphrased documents became ever more sophisticated, several techniques were applied to the same text passages, for example synonym and generalization.

\subsection{The basic/simple paraphrase sub-corpus}
\label{basi_simple}

To start with, the five documents of the basic paraphrase corpus were created.
The first document was obtained by applying some of the rather basic changes mentioned above to the original Wikipedia article.
Subsequent documents were built on the text of documents that had been created in a previous step.
Sometimes a new phrase was added or deleted in accordance with the structural rules.
Then further changes were made, such that step by step the list of techniques mentioned above was elaborated.
To check for the effect of permutations at least one of the paraphrased versions was a simple permutation of a previous one, may be plus one additional phrase or minus one deleted phrase.
It is worthwhile to mention that the last document of the basic paraphrase corpus made use of almost all of the techniques that had been developed, but it respected all the structural rules associated with the basic corpus, in particular no exchange of segments and minimal or no change in the length of the document.

\subsection{The complex paraphrase sub-corpus}
\label{complex}

The first document of the complex paraphrase corpus was created in order to evaluate the effect of bigger structural changes.
Therefore several phrases were added to and deleted from a document of the basic paraphrase corpus, one phrase was merged and one split.
The sequence of phrases was not changed.

The second document was based on the first one, with more splits, merges and exchanges of segments between phrases.
Furthermore the order of the phrases was permuted. The level of paraphrase was also raised.
Among others, all numbers were written in textual form or vice versa, geographic and temporal references were generalized and the use of synonyms and definitions of terms increased.

The third document used one of the more sophisticated documents of the basic paraphrase corpus as a starting point.
The level of paraphrase was increased by using techniques from the list above more frequently.
Phrases were added, removed, split and merged, but no segments moved between phrases.

The fourth document took the most sophisticated document of the basic paraphrase corpus as a starting point.
A special effort was then made to apply all the techniques mentioned so far to the maximum, especially generalization of personal, geographic and other entities.
New paraphrases were devised that had not been used in other documents of either of the sub-corpora.
The maximum allowed number of phrases was added and deleted, several phrases merged and split.
In addition, several segments were moved between phrases. Finally, the phrases were reordered in compliance with the comprehensibility requirement.
In essence, the fourth document is the most elaborately paraphrased of all the documents in the entire corpus.

As almost all the options had been applied to a certain degree, the fifth document was compiled by selecting sets of phrases from several previously mentioned documents, then deleting the maximum allowed number, adding 3 new phrases, merging some and moving segments where appropriate.

\section{Evaluation aids}
\label{evaluation}

Similarity between the original document and each of the paraphrased versions was calculated by applying the techniques and formulas described in the previous sections.
Basically, the algorithms tried to determine the most similar phrases in the original to any of the phrases in the paraphrased document and vice versa.
To be able to assess the precision of those algorithms' predictions, a simple method to specify the relationship between the original and the paraphrased document was devised.

The mapping between phrases of the original document and the phrase or phrases in the paraphrased documents was noted, transformed into a predefined format and then stored into a small file specific to each document.
This file would contain one line for each phrase in the original.
The line begins with the number of the phrase in the original document, then a colon and then the number of the phrase in the paraphrased document to which the original was rewritten.
The following examples illustrate this mapping technique. Please also note that ``phrase X of the original is found in phrase Y of the paraphrased document'' is shorthand for ``all or part of the meaning of phrase X after much paraphrasing can be found in phrase Y''.

\begin{itemize}
 \item ``0 : 29'' indicates that phrase 0 of the original is found in phrase 29 of the paraphrased document.
 \item ``2:'' indicates that phrase 2 of the original has been removed.
 \item ``3: 1'' together with ``4 :1'' indicate that phrases 3 and 4 from the original have been merged into phrase 1 of the paraphrased document.
 \item ``13 : 11,12'' indicates that phrase 13 from the original has been split and may be found in phrases 11 and 12 of the paraphrased document.
 \item ``5: 5,6'' together with ``6: 5,6'' indicate that segment of phrases 5 and 6 from the original have been exchanged.
 \item If there is a phrase in the paraphrased document the number of which is to be found nowhere to the right of the colon in the file specifying the mapping, that means this phrase has been added to the original.
\end{itemize}

\section{Exploitation of German corpus}
\label{exploitation}

We assessed the average values of the simple similarity measures calculated on the 15 texts of the German corpus (five texts of the basic paraphrase sub-corpus, five texts of the complex paraphrase sub-corpus and five texts not related with the source text).
Except for Euclidean distance, Manhattan distance and Jaro-Winkler (JW), the measures were normalized to the range $[0,1]$.


Values close to 0 indicate high proximity between the source text and the suspicious text 
(although for the Euclidean, Manhattan and JW distance higher values indicate more difference between both texts).

\begin{figure}[!t]
\centering
\includegraphics[ width=0.8  \textwidth ]{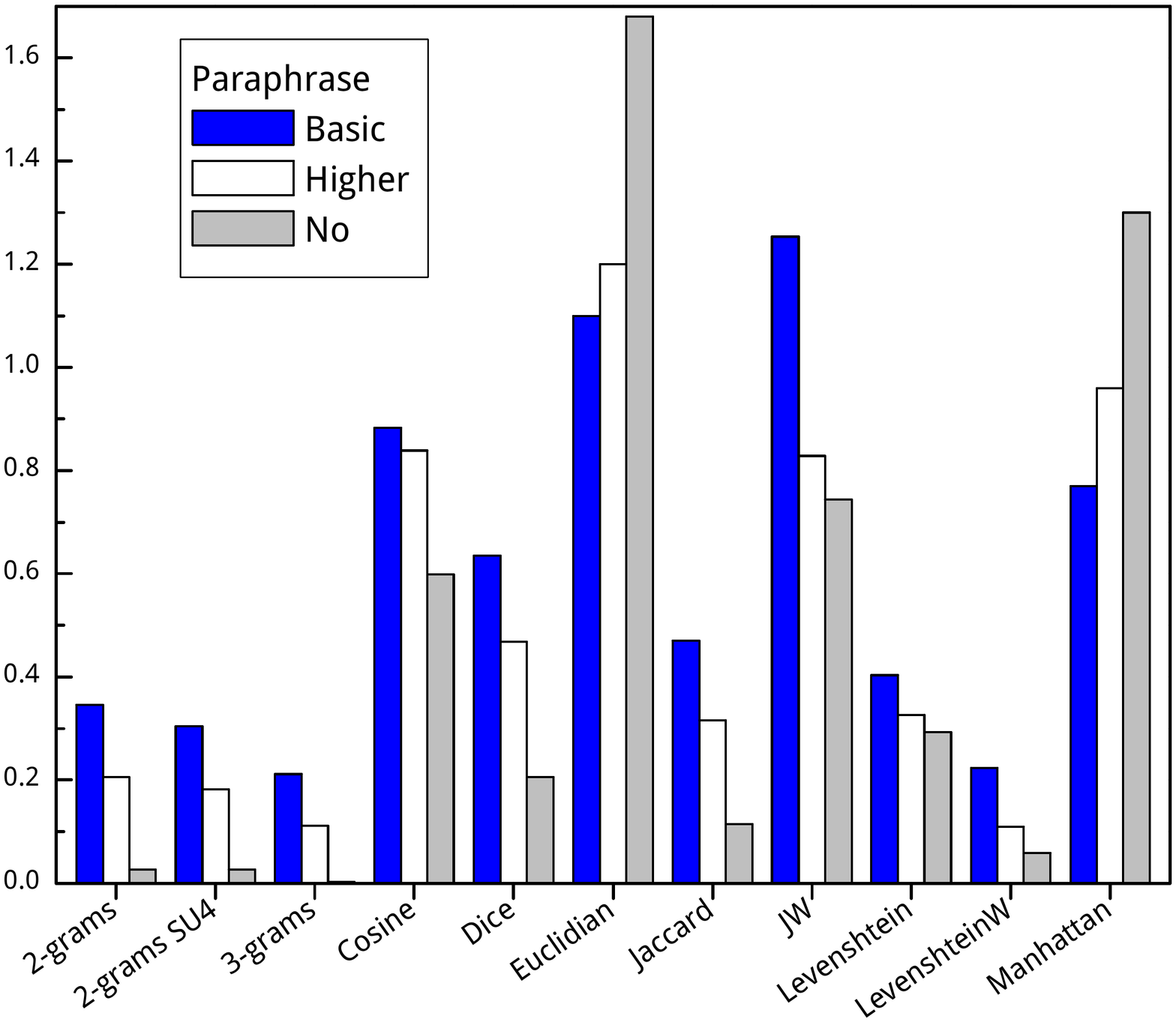}
\caption{Text similarity measures calculated over the German corpus: not paraphrase.}
\label{notparaphrase}
\end{figure}

As can be seen in Figure \ref{notparaphrase}, the highest similarity scores correspond to paraphrases of the basic level, following the higher level and finally the lowest score for no paraphrases (but inverted scores for Euclidean and Manhattan distances).

We also calculated the Pearson correlation factor \cite{mihalcea06} among all the similarity measures to assess the score among overall simple measures.
Table \ref{simil} shows a strong correlation as was expected.

\begin{table}
\begin{center}
\begin{tabular}{|l|c|c|c|c|}
  \hline
  \bf Similarity    & Low     & High    & Not     & Pearson  \\
  \hline
  2-grams       & 0.34565 & 0.20559 & 0.02875 & 0.97716 \\
  2-grams SU4   & 0.30396 & 0.18248 & 0.02971 & 0.97692 \\
  3-grams       & 0.21197 & 0.11161 & 0.00178 & 0.96765 \\
  Cosine        & 0.88245 & 0.83879 & 0.63233 & 0.99693 \\
  Dice          & 0.63473 & 0.46795 & 0.22470 & 0.98494 \\
  Euclidean     & 1.07344 & 1.22539 & 1.67582 & -1      \\
  Jaccard       & 0.46983 & 0.31623 & 0.12709 & 0.97562 \\
  JW            & 1.25345 & 0.82801 & 0.73783 & 0.80201 \\
  Levenshtein   & 0.40322 & 0.32613 & 0.28818 & 0.88927 \\
  Levenshtein W & 0.22375 & 0.10933 & 0.05669 & 0.88169 \\
  Manhattan     & 0.78558 & 0.95836 & 1.29115 & -0.995  \\ \hline
\end{tabular}
\caption{Correlation of similarity measures.}
\end{center}
\label{simil}

\end{table}

\section{Conclusions}
\label{conclusions}

We presented a new German corpus for paraphrasing detection.
It features two particular characteristics as compared to current corpora, i.e. a set of aligned text pairs at paraphrase level with a file of references, and three levels of paraphrase for each document (high, low and no paraphrase).

The mapping between the source text and the paraphrased one lets even persons that do not speak German study the corpus for similarity detection, thereby making it language independent.
Furthermore, the level of paraphrase allows to refine the algorithms for paraphrasing detection, in order to determine the degree of paraphrase more precisely.
 
The application of the simple measures on our corpus and of the Pearson correlation factor overall measures let us see, as expected, how the similarity score increases in direct proportion to the degree of paraphrase.

We are still incorporating new texts into the German corpus.
Besides, we are preparing two other corpora, a Spanish and a French corpus, using the same protocol as the one presented in this paper.
Meanwhile, the German corpus is available online at the website: \url{http://simtex.talne.eu}

\section*{Acknowledgments}
We acknowledge the Mexico's {\sl Consejo Nacional de Ciencia y Tecnolog\'ia} (Conacyt) grant number 178248 and Project  UNAM-DGAPA-PAPIIT number IN\-400312. 

\bibliographystyle{splncs}
\bibliography{paper}

\end{document}